# Multi-Modality Abdominal Multi-Organ Segmentation with Deep Supervised 3D Segmentation Model

Satoshi Kondo[1][0000-0002-4941-4920] and Satoshi Kasai[2]

[1] Muroran Institute of Technology, Hokkaido, Japan
[2] Niigata University of Health and Welfare, Niigata, Japan
`kondo@mmm.muroran-it.ac.jp`

**Abstract.** To promote the development of medical image segmentation technology, AMOS, a large-scale abdominal multi-organ dataset for versatile medical image segmentation, is provided and AMOS 2022 challenge is held by using the dataset. In this report, we present our solution for the AMOS 2022 challenge. We employ residual U-Net with deep super vision as our base model. The experimental results show that the mean scores of Dice similarity coefficient and normalized surface dice are 0.8504 and 0.8476 for CT only task and CT/MRI task, respectively.

## 1 Introduction

Whole abdominal organ segmentation plays an important role in diagnosing abdomen lesions, radiotherapy, and follow-up. The recent success of deep learning methods applied for abdominal multi-organ segmentation exposes the lack of large-scale comprehensive datasets for developing and comparing these methods.

To promote the development of medical image segmentation technology, AMOS, a large-scale abdominal multi-organ dataset for versatile medical image segmentation, is provided [1]. It provides 500 CT and 100 MRI scans collected from multi-center, multi-vendor, multi-modality, multi-phase, multi-disease patients, each with voxel-level annotations of 15 abdominal organs. And multi-modality Abdominal Multi-Organ Segmentation challenge 2022 (AMOS 2022) is held by using the above dataset [2]. AMOS 2022 contains two tasks as followings.

a) Task 1 - Segmentation of abdominal organs (CT only): It aims to comprehensively evaluate the performance of different segmentation methods across large-scale and great diversity CT scans, a total of 500 cases.

b) Task 2 - Segmentation of abdominal organs (CT & MRI): It extends the image modality target of task 1 to the MRI modality. Under a multi-modality setting, a single algorithm is required to segment abdominal organs from both CT and MRI. Specifically, additional 100 MRI scans with the same type of annotation will be provided.

In this report, we present our solution to the AMOS 2022 challenge. We employ residual U-Net with deep super vision as our base model. We use the same model architecture and the training procedure except the normalization parameters for tasks 1 and 2.

## 2    Proposed Method

We use 3D segmentation network as our base model. Our base model is residual U-Net with deep super vision [3]. An input volume is resampled in [1mm, 1mm, 2mm] for x, y and z direction, respectively, at first. CT volumes are normalized with clipping. The minimum and maximum values are -100 and 250, respectively, for the clipping. MRI volumes are normalized by subtracting the mean value of the volume and dividing the standard deviation value of the volume. In the training phase, we randomly sample 3D patches from the input volumes. The size of a 3D patch is 128 x 128 x 64 voxels. The ratio of positive and negative patches in the sampling for one input volume is 1:1. We apply intensity shift within 5 % for augmentation.

The loss function is weighted summation of Dice loss and cross entropy loss. The background pixels are excluded in the Dice loss. The weights for Dice loss and cross entropy loss are 1.0 and 0.5, respectively. We also employ deep super vision for loss calculation. Intermediate outputs from several layers in the decoder of the model are up-sampled, loss value is calculated for each up-sampled output, and then the loss values are aggregated. The number of layers used in the deep super vision is three.

The optimizer is Adam [3] and the learning rate changes with cosine annealing. The initial learning rate is 0.001. The number of epoch is 300. The model taking the lowest loss value for the validation dataset is selected as the final model.

We train multiple models. Each model is trained independently using different combinations of training and validate datasets, and the inference results are obtained by ensemble of the outputs from the models. The final likelihood score is obtained by averaging the likelihood scores from the models. We use five models in our experiments.

We use the same training procedures for Task 1 and Task 2. The only difference in these tasks is the normalization method as mentioned earlier, i.e. we use different normalization parameters for CT and MRI volumes

## 3    Experiments

We evaluated our method with the evaluation system provided by the organizers of AMOS 2022. For the first phase (validation phase) in the evaluation, 100 volumes for Task 1 and 120 volumes for Task 2 are used. The evaluation metrics are Dice Similarity Coefficient (DSC), and Normalized Surface Dice (NSD).

The results of our submission in Task 1 are DSC is 0.8936 and NSD is 0.8071. The results of our submission in Task 1 are DSC is 0.8880 and NSD is 0.8071.

## 4  Conclusions

In this report, we presented our solution for the AMOS 2022 challenge. We employ residual U-Net with deep super vision as our base model. The experimental results show that the mean scores of DSC and NSD are 0.8504 and 0.8476 for tasks 1 and 2, respectively.